\documentclass[twocolumn,showpacs,preprintnumbers,amsmath,amssymb]{revtex4}

\usepackage{graphicx}
\usepackage{dcolumn}
\usepackage{bm}

\DeclareTextSymbol{\degre}{T1}{6}
\DeclareTextSymbol{\degre}{OT1}{23}

\begin{document}


\title{Magneto-transport in high g-factor, low-density
two-dimensional electron systems confined in
In$_{0.75}$Ga$_{0.25}$As/In$_{0.75}$Al$_{0.25}$As quantum wells}

\author{W. Desrat}
\author{F. Giazotto}
\author{V. Pellegrini}
\author{F. Beltram}
\affiliation{NEST-INFM and Scuola Normale Superiore, Piazza dei
Cavalieri 7, I-56126 Pisa, Italy}

\author{F. Capotondi}
\altaffiliation[Also at ]{Dipartimento di Fisica, Universit\'a di
Modena e Reggio Emilia, I-43100 Modena, Italy}

\author{G. Biasiol}
\author{L. Sorba}
\altaffiliation[Also at ]{Dipartimento di Fisica, Universit\'a di
Modena e Reggio Emilia, I-43100 Modena, Italy}
\affiliation{NEST-INFM and Laboratorio Nazionale TASC-INFM, Area
Science Park, I-34012 Trieste, Italy}

\author{D.K. Maude}
\affiliation{Grenoble High Magnetic Field Laboratory, MPI-FKF and
CNRS, BP 166, 38042 Grenoble Cedex 9, France}

\date{\today}

\begin{abstract}

We report magneto-transport measurements on high-mobility
two-dimensional electron systems (2DESs) confined in
In$_{0.75}$Ga$_{0.25}$As/In$_{0.75}$Al$_{0.25}$As single quantum
wells. Several quantum Hall states are observed in a wide range of
temperatures and electron densities, the latter controlled by a
gate voltage down to values of $1\times10^{11}$ cm$^{-2}$. A
tilted-field configuration is used to induce Landau level
crossings and magnetic transitions between quantum Hall states
with different spin polarizations. A large filling factor
dependent effective electronic g-factor is determined by the
coincidence method and cyclotron resonance measurements. From
these measurements the change in exchange-correlation energy at
the magnetic transition is deduced. These results demonstrate the
impact of many-body effects in tilted-field magneto-transport of
high-mobility 2DESs confined in
In$_{0.75}$Ga$_{0.25}$As/In$_{0.75}$Al$_{0.25}$As quantum wells.
The large tunability of electron density and effective g-factor,
in addition, make this material system a promising candidate for
the observation of a large variety of spin-related phenomena.
\end{abstract}


\maketitle

\section{Introduction}
A number of spintronic applications would be experimentally
accessible thanks to the availability of high-mobility
two-dimensional electron systems (2DESs) confined in
In$_x$Ga$_{1-x}$As layers with $x\ge 0.75$. This stems from three
different factors. First, metallic contacts on In$_x$Ga$_{1-x}$As
with such high Indium concentration are characterized by the
absence of a Schottky barrier at the interface. This naturally
provides a highly transmissive junction paving the way to the
observation of Andreev reflection phenomena in superconductor/2DES
hybrid systems and their study in the regime of the quantum Hall
effect \cite{Jakob00,Uhlisch00}. Second, the bare g-factor is
expected to be very large. For instance the bare g-factor measured
by electron spin resonance was found at around -5 in
In$_{0.53}$Ga$_{0.47}$As \cite{Dobers89} while effective g-factors
$g^*$ (renormalized by electron-electron interactions) with
absolute values between $14$ and $28$ were recently suggested from
magneto-transport data \cite{Kita01,Gui00}. The large resulting
Zeeman spin splitting under the application of weak magnetic
fields makes this semiconductor system a promising candidate for
spin-valve mesoscopic devices working at relatively high
temperatures \cite{Gilbert00}. Third, the large Rashba coupling
constant that characterizes asymmetric In$_x$Ga$_{1-x}$As layers
\cite{Nitta97} can be used for the control of spin-dependent
transport and for the creation of spin-Hall currents
\cite{Nitta97,Jungwirth03}. For all of these investigations, the
combination of high-mobility and relatively low electron density
is a crucial requirement as well as the gate-voltage control of
the various parameters mentioned above.

In addition, these systems can be of much interest for the study
of magnetic phenomena associated to the crossing of Landau levels
with opposite spin polarizations. These crossings can be induced
in a tilted-field configuration since the ratio between the Zeeman
energy ($E_Z$) and the cyclotron energy ($E_C$) increases at
larger values of the tilt angle. For a large enough tilt angle
this ratio reaches unity and a crossing between Landau levels of
neighboring orbital indices and opposite spins is expected. It was
shown, however, that this magnetic transition occurs before the
coincidence between the single-particle energy levels and is
induced by exchange-correlation effects \cite{Giuliani85,Koch93}.
These transitions are triggered by the behavior of the long
wave-vector limit of the spin gap (that includes many-body
contributions) and therefore can be also used to determine the
exchange-enhanced electronic g-factor ($g^*$) \cite{Nicholas88}.
In the past, the experimental investigation of this class of
many-body-driven magnetic transitions was mainly carried out on
high-mobility 2DESs confined in GaAs/AlGaAs heterostructures. The
exploitation of heterostructures based on In$_x$Ga$_{1-x}$As would
allow to investigate the influence of both effective mass and
g-factor. In addition, the presence of a large Rashba coupling
could yield new coherent Landau level configurations close to
coincidence as predicted in Refs.\cite{Falko92,Falko93}. Recent
tilted-field experiments performed on 2DESs confined in InAs/InSb,
however, showed no evidence of electron-electron interaction
effects presumably due to a combination of large electron density
(above $6\times 10^{11}$ cm${-2}$) and relatively low mobility
\cite{Brosig00}.

\begin{figure}
\includegraphics[width=7.6cm]{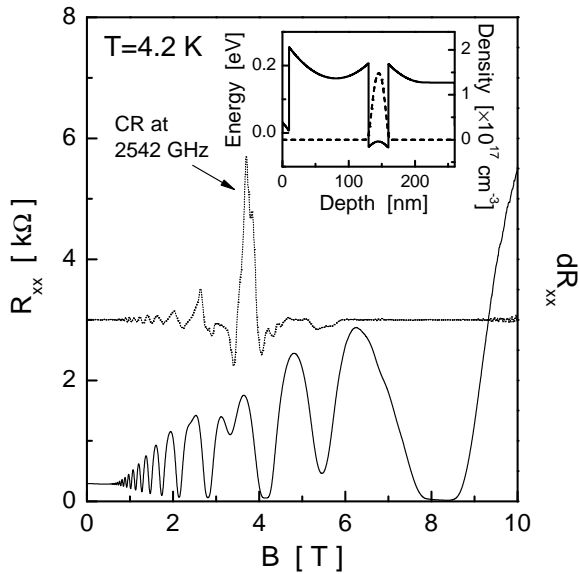}
\caption{\label{figure1} Longitudinal resistance $R_{xx}$ (solid
curve, vertical left axis) and resistance change $dR_{xx}$ (dotted
curve) after far-infrared radiation at $2.542$ THz as a function
of magnetic field at $T=4.2$ K. The inset shows the
conduction-band diagram of the heterostructure and charge
distribution obtained from 1D Poisson-Schr\"odinger simulation
assuming a uniform distribution of deep donor states in the
barriers.}
\end{figure}

Here we report the realization of 2DESs with mobilities well above
$10^5$ cm$^2$/Vs confined in
In$_{0.75}$Ga$_{0.25}$As/In$_{0.75}$Al$_{0.25}$As single quantum
wells. These electron systems display clear quantum Hall states in
a large range of temperatures and with electron densities down to
$n_s=1\times 10^{11}$ cm$^{-2}$, the latter controlled by a gate
voltage. In this paper we also report the magneto-transport
analysis in a tilted magnetic-field configuration under magnetic
fields up to $23$ T and at millikelvin temperatures. This analysis
shows crossing of spin-split Landau levels at several even-integer
filling factors. These results are used to determine the
exchange-enhanced "effective" electronic g-factor. Values of $g^*$
are carefully deduced from the coincidence method with electron
mass determined by cyclotron resonance and they are found to be
$\nu$-dependent. Measurements performed on a gated sample confirm
this behavior for different densities. The enhanced g-factor
decreases and saturates at high filling factors in agreement with
the reduction of the exchange energy. From these data the change
in the exchange-correlation energy at the magnetic transition
between Landau states with opposite spins is also measured and is
found to vary linearly as a function of the perpendicular
component of the magnetic field. These results demonstrate the
impact of electron-electron interactions on magnetic transitions
in In$_{0.75}$Ga$_{0.25}$As/In$_{0.75}$Al$_{0.25}$As quantum
wells.

\section{Gate-voltage control of electron density}
Samples analyzed in this work consist of an unintentionally doped
$30$ nm-wide In$_{0.75}$Ga$_{0.25}$As/In$_{0.75}$Al$_{0.25}$As
quantum well metamorphically grown by molecular beam epitaxy on a
(001) GaAs substrate. An In$_x$Al$_{1-x}$As step graded buffer
with $x$ ranging from 0.15 to 0.85 was inserted between the
substrate and the 2DES to achieve almost complete strain
relaxation at the quantum-well region. In these nominally undoped
structures, a relatively low carrier concentration close to
$3\times10^{11}$ cm$^{-2}$ can be achieved and this confirms that
there exists an intrinsic source of free carriers
\cite{Schapers98}. The origin of the doping is, however, still
unclear. Preliminary results indicate the role of deep-level donor
states lying in the conduction band discontinuity
\cite{Capotondi03}. The inset of Fig.\ref{figure1} shows the
conduction band diagram and the charge distribution obtained from
a Poisson-Schr\"odinger simulation \cite{Snider} assuming a
uniform distribution of deep levels with activation energy of 0.12
eV. The latter was measured by photo-induced current transient
spectroscopy as reported elsewhere \cite{Capotondi03}. The main
panel of Fig.\ref{figure1} displays the longitudinal resistance
$R_{xx}$ (solid curve, vertical left axis) as a function of
perpendicular magnetic field at $T=4.2$ K. The zero-resistance
regions which are well defined at even filling factors indicate
the absence of parallel conduction and are associated to a
quantized Hall resistance. From this data we extract a density of
$n_s=3.1\times10^{11}$ cm$^{-2}$ with a mobility above
$\mu=2\times10^{5}$ cm$^2$/Vs (electron mean free path of the
order of 1.8 $\mu$m). Cyclotron resonance is shown on
Fig.\ref{figure1} (dotted curve, right vertical axis) and
manifests as a peak in the longitudinal resistance change
$dR_{xx}$ after far-infrared radiation at $2.542$ THz as a
function of magnetic field. The absorption peak is observed at
$B=3.57$ T which yields the effective mass of the conduction
electrons equal to $m^*=0.039\pm0.002\;m_0$, where $m_0$ is the
free electron mass.

\begin{figure}[]
\includegraphics[width=7.6cm]{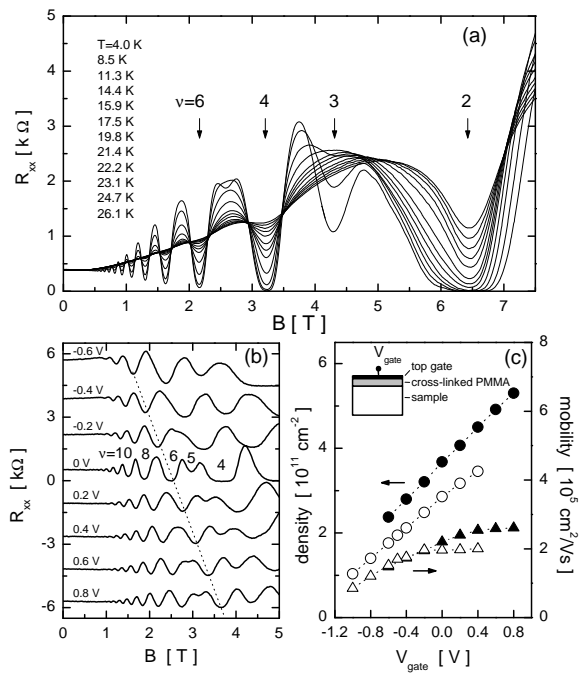}
\caption{\label{figure3} (a) Longitudinal resistance $R_{xx}$
versus magnetic field $B$ as a function of temperature. Data
correspond to sample with electron density $n_s=3.1\times 10^{11}$
cm$^{-2}$ and mobility $\mu=2.1\times 10^5$ cm$^2$/Vs. (b)
$R_{xx}$ versus magnetic field for different values of the gate
voltage $V_{gate}$ at $T=360$~mK. (c) Gate voltage dependence of
electron density (left vertical axis) and mobility (right vertical
axis) for two different samples at $T=360$~mK. The sketch of the
sample with gate is shown in the upper part of the panel.}
\end{figure}

\begin{figure*}[t]
\includegraphics[width=15.2cm]{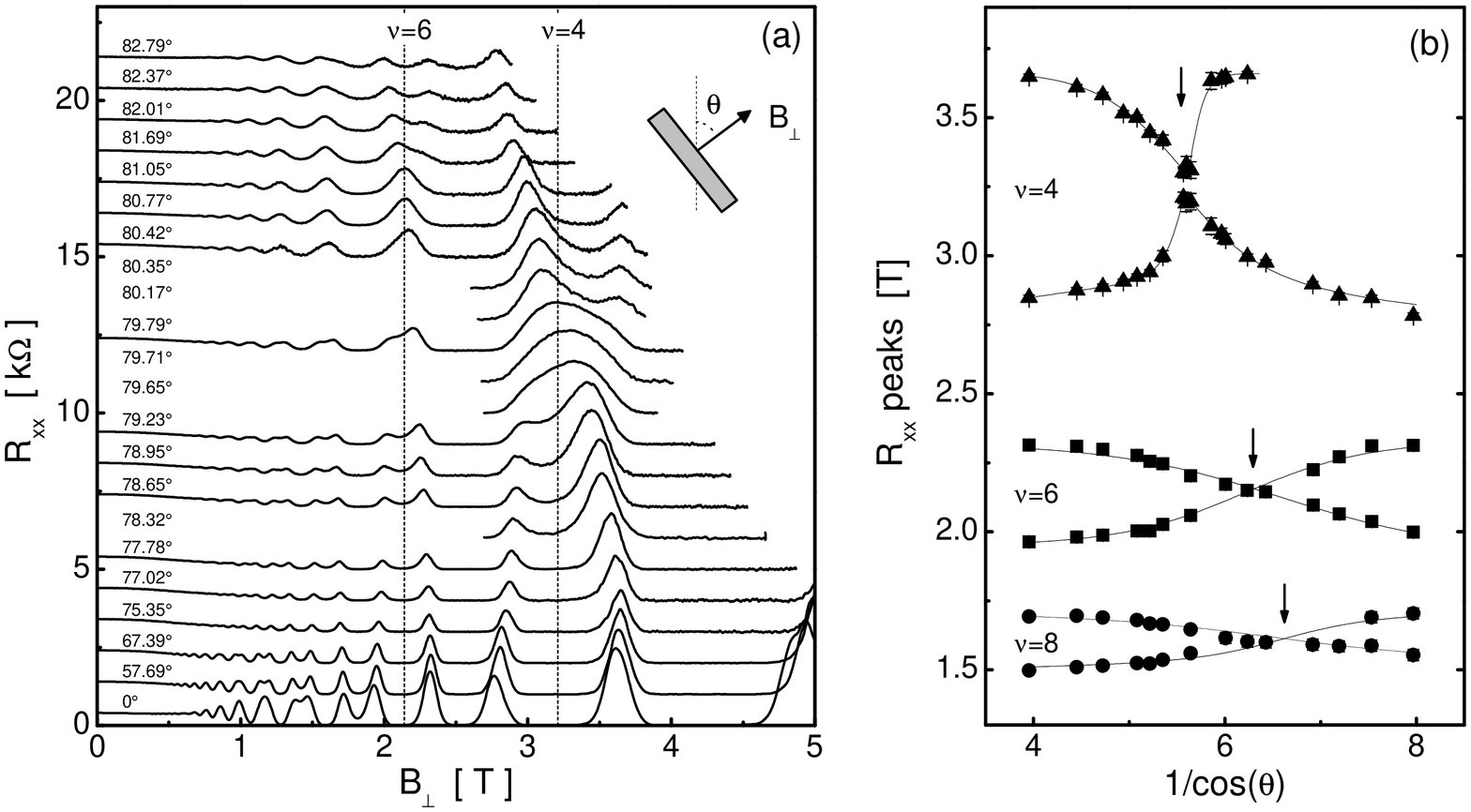}
\caption{\label{figure2} (a) Perpendicular magnetic field
dependence of the longitudinal resistance for different tilt
angles $\theta $. The latter were accurately obtained from the
Hall voltage. The curves are shifted for clarity; the inset shows
a sketch of the tilted-field geometry. (b) Perpendicular magnetic
field position of the $R_{xx}$ peaks vs $1/\cos(\theta)$. The thin
lines are guides for the eye; the arrows indicate the angles
($\theta= 79.68^\circ$, $80.89^\circ$ and $81.67^\circ$) at which
the first coincidence of Landau levels occurs at even-integer
filling factors ($\nu=4,6$ and 8 respectively).}
\end{figure*}

The impact of the large g-factor and low electron mass that
distinguishes this system from 2DESs confined in GaAs quantum
wells is further shown in the temperature-dependent
magneto-transport data displayed in Fig.\ref{figure3}(a). The data
demonstrates, in particular, that the odd-integer quantum Hall
minima at $\nu=3$ and $5$ associated to the Landau level spin gap
are observable at $T=4.2$~K (up to $T\approx 11$~K for the $\nu=3$
state) indicating the presence of robust spin-polarization at
relatively low magnetic fields. An important aspect for spintronic
applications in this material system stems from the possibility to
reach low electron density and to control the asymmetry of the
confining potential (i.e. Rashba coupling as shown in
Ref.\cite{Schapers98}) still maintaining large values of the
electron mobility. In this case several recently predicted
spin-related phenomena can be studied. These include the
generation of an intrinsic spin-Hall current and the control of
its value for densities typically below $n_s= 1-1.5\times 10^{11}$
cm$^{-2}$ \cite{Jungwirth03}. Such low densities and their gate
voltage control have never been observed in high-mobility 2DESs
confined in InGaAs quantum wells with very large In concentration.
Figures \ref{figure3}(b) and (c) demonstrate the successful
achievement of these desired properties. The gate voltage control
of the electron density is obtained by the application of a
voltage difference between a top Ti/Au gate deposited onto a
cross-linked PMMA thin layer (thickness of the order of 50 nm) and
the substrate. Figure \ref{figure3}(b), in particular, reports the
longitudinal resistivities without any parallel conduction at
$T=4.2$~K as a function of the gate voltage while
Fig.\ref{figure3}(c) displays the density $n_s$ (left axis) and
mobility $\mu$ (right axis) as a function of gate voltage for two
different devices. A good control on the value of $n_s$ down to a
value of 1$\times$ 10$^{11}$ cm$^{-2}$ can be obtained despite the
uniform distribution of doping states and without altering
significantly the electron mobility.

It must be noted, in addition, that the generated electric field
introduces an asymmetry in the confining potential thus enhancing
the value of the Rashba coupling \cite{Nitta97}. In this
relatively low-density In$_{0.75}$Ga$_{0.25}$As quantum well only
a limited number of Shubnikov-de Haas oscillations are detected at
relatively low magnetic fields which inhibits the conventional
determination of the Rashba coupling constant of the
heterostructure. Previous studies, in fact, were conducted on
samples having densities approaching $1\times10^{12}$ cm$^{-2}$ or
above \cite{Nitta97,Schapers98,Koga02}. It can be noted in
Fig.\ref{figure3}(b), however, that an anomalous behavior
characterizes the longitudinal resistance close to filling factor
$\nu=4$. The $\nu=4$ minimum, in fact, tends to disappear at
positive non-zero values of the gate voltage and this might be
interpreted as precursor of the beating pattern associated to the
enhancement of the Rashba coupling for a large quantum well
asymmetry generated by the gate voltage \cite{Schapers98,Koga02}.
The analysis of the transverse resistance (data not shown) also
strongly suggests a significant Rashba coupling consistent with
the theoretical prediction of Ref.\cite{Wang03}.

\section{Enhanced g-factor and magnetic transitions in tilted fields}

Having demonstrated good magneto-transport properties and their
gate-voltage control we now turn to the study of the evolution of
the magneto-transport in a tilted-field configuration at ultra-low
temperatures. As shown below these data allow to determine the
effective (exchange-enhanced) g-factor and to identify the impact
of many-body effects on the magnetic transitions induced by the
tilted field. For these measurements the sample was mounted on a
rotation probe and placed in the mixing chamber of a dilution
fridge. The maximum available magnetic field is $23$ T and the
bath temperature is $30$ mK. The longitudinal and Hall resistances
are measured on a Hall bar via a lock-in setup with a low ac
current ($I=20 $ nA, $f=17$ Hz). Figure \ref{figure2}(a) shows a
waterfall graph of the longitudinal resistance $R_{xx}$ as a
function of the perpendicular magnetic field for different tilt
angles $\theta$ (see inset of Fig.\ref{figure2}(a)) at $T=30$ mK.
At such low temperatures and at $\theta=0^\circ$ several
dissipationless QH regions at both even and odd filling factors
are clearly seen from $\nu =1$ (not shown) up to the filling
factor $\nu=12$. As the tilt angle increases, the width of these
zero-resistance regions decreases for even-integer filling factors
($\nu=4,6,8,10,12$) and increases for odd-integer filling factors
($\nu=3,5,7$). This behavior is in agreement with the evolution of
the respective energy gaps in the density of states as a function
of $\theta$ linked to the increase of the ratio $E_{Z}/E_{C}$. For
angles larger than $81^\circ$, however, the inverse dependence
takes place, i.e. the even filling factor gaps become larger and
those at odd fillings are reduced. For intermediate tilt angles,
the zero-resistance region corresponding to odd-integer filling
factors ($\nu=5,7$) saturates to its maximum width while the
$R_{xx}$ minima of the even-integer filling factors disappear and
are replaced by maxima. This behavior signals the transition
between QH phases with different spin polarizations and can be
used to determine $g^*$.

Figure \ref{figure2}(b) displays the perpendicular magnetic field
position of the $R_{xx}$ peaks of the Shubnikov-de Haas
oscillations as a function of the inverse cosine of the tilt angle
$\theta$. These peak positions were estimated from the derivative
curves $dR_{xx}/dB$. Coincidences between spin-split Landau
levels, indicated by arrows, occur at different values of
$1/\cos\theta _{c}$ (5.58, 6.32 and 6.62) for different
even-integer filling factors ($\nu=4,\;6$ and 8 respectively).
Using these coincidence angles and the measured effective mass, we
can determine the exchange-enhanced g-factor $\vert g^*\vert
=2m_0\cos(\theta_c)/m^*$ \cite{Nicholas88}. We obtain $\vert
g^*\vert=9.2,8.1,7.7$ for $\nu=4,6,8$, respectively. The reduction
of $\vert g^*\vert$ for larger filling factors was already
reported in other material systems and is interpreted as due to
the reduction of electron-electron interaction effects
\cite{Ando74}.

\begin{figure*}[]
\includegraphics[width=15.2cm]{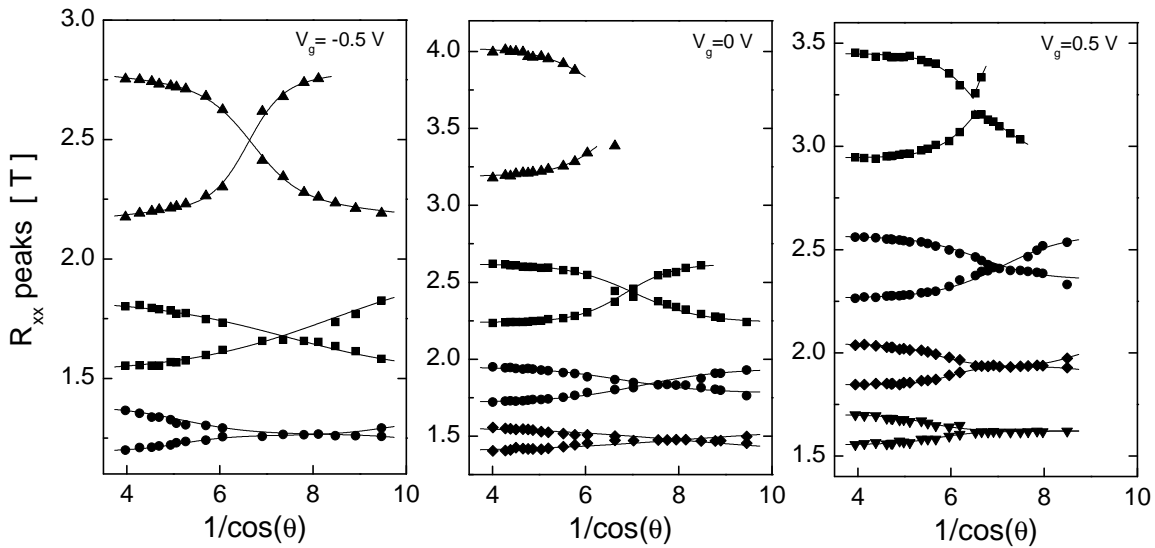}
\caption{\label{figure4} Perpendicular magnetic field position of
the $R_{xx}$ peaks as a function of $1/\cos(\theta)$ for a
In$_{0.75}$Ga$_{0.25}$As/In$_{0.75}$Al$_{0.25}$As quantum well
with a density of $n_s=3.7\times10^{11}$~cm$^{-2}$ and a mobility
of $\mu=7.5\times10^4$~cm$^2$/Vs. The graphs (from left to right)
have been obtained at three different gate voltages :
$V_g=-0.5$~V,$0$~V and $+0.5$~V respectively. Here data is
presented for filling factors $\nu=4$ (up triangles), 6 (squares),
8 (disks), 10 (diamonds) and 12 (down triangles). The thin lines
are guides for the eye.}
\end{figure*}

A similar tilted-field experiment was performed on a second
In$_{0.75}$Ga$_{0.25}$As/In$_{0.75}$Al$_{0.25}$As quantum well
with a density of $n_s=3.7\times10^{11}$~cm$^{-2}$ and a mobility
of $\mu=7.5$~m$^2$/Vs measured at $T=1.4$~K. A gate on top of this
second sample allows to vary the electron density. Figure
\ref{figure4} shows the perpendicular magnetic field position of
the $R_{xx}$ peaks as a function of the inverse cosine of the tilt
angle for three different gate voltages. The densities
(mobilities) are 2.55 (5.2) and 4.9 (10) at $V_g=-0.5$~V and
$V_g=+0.5$~V respectively, in units of $10^{11}$~cm$^{-2}$ and
m$^2$/Vs. In all these configurations the magnetic transitions
between Landau levels with opposite spins occur at different tilt
angles for different even filling factors thus confirming the
impact of the many-body corrections in the material system here of
interest.

\begin{figure}[]
\includegraphics[width=7.6cm]{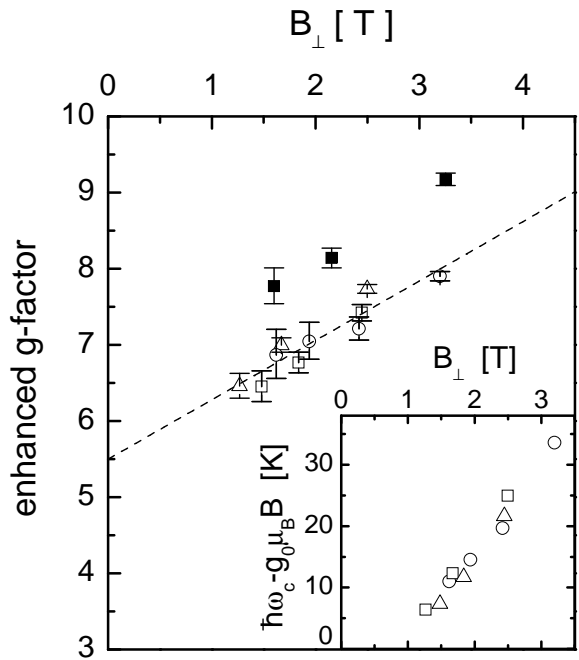}
\caption{\label{figure5} Enhanced electronic g-factor (defined as
$2m_0\cos(\theta_c)/m^*$) as a function of perpendicular magnetic
field for both samples presented in Fig.\ref{figure2}b (solid
squares) and Fig.\ref{figure4} (open symbols). The latter provides
three sets of values obtained at different gate voltages :
$V_g=-0.5$~V (open circles), $V_g=0$~V (open squares) and
$V_g=+0.5$~V (open triangles). The dashed line shows a linear
dependence of $g^*$ vs $B_\perp$ with a zero field value equal to
5.5. In inset, energy difference, $\hbar\omega_c-g_0\mu_BB$, as a
function of $B_\perp$.}
\end{figure}

The observed coincidence angles yield  values for the enhanced
g-factors plotted on Fig.\ref{figure5} as a function of the
perpendicular component of the magnetic field $B_{\perp}$. The
enhanced g-factor decreases for lower perpendicular magnetic field
(or higher filling factors) and lower mobilities. Having in mind
that the long wavevector limit of the spin gap (relevant in
magneto-transport) is the sum of the bare Zeeman energy
$g_0\mu_BB$ and the exchange energy proportional to
$e^2/4\pi\varepsilon\ell_B$, it is tempting to describe the
behavior of $g^*(B_{\perp})$ by a $\sqrt{B_{\perp}}$ dependence.
However, effects associated to mixing of Landau levels, disorder,
finite-layer thickness and Coulomb correlation among electrons
could play a role especially in the limit of low density. The
behavior shown in Fig.\ref{figure5} in fact, shows a linear rather
than a square root dependence on $B_{\perp}$. Additionally, the
data enlighten the fact that the many-body contribution is
enhanced when the mobility increases and becomes negligible at
high filling factors when the enhanced g-factor saturates to the
bare g-factor $g_0$, which is estimated of the order of $5.5$ (in
absolute value). We note that the estimated value of $g_0$ is
small compared to the theoretical one of bulk
In$_{0.75}$Ga$_{0.25}$As equal to $\vert g_0\vert=8.9$. However,
several recent papers also reported g-factor values strongly
reduced compared to the theoretical ones. For instance, bare
g-factors of -13, -8.7 or -6 have been reported in InAs quantum
wells \cite{Brosig00,Hu03,Moller03} which are far from the bulk
value of around -15. It must be stressed that in the present
experiments the orbital effect of the in-plane field becomes
important at large tilt angles and may contribute to the reduction
of $g^*$. It is well known, in fact, that the measurement of the
electronic enhanced g-factor, and consequently of the bare
g-factor, from magneto-transport measurements is not
straightforward as it depends on the polarization of the system,
in other words on the position of the Fermi energy in the density
of states \cite{Nicholas83}.

Now, with the value of the bare g-factor of $\vert g_0\vert=5.5$,
it is possible to compute the energy difference, $\hbar
eB_\perp/m^*-g_0\mu_BB_\perp/\cos(\theta_c)$, for all the tilt
angles $\theta_c$ at which the transition is observed. This is
performed for all filling factors and all gate voltages. In the
single-electron picture, this difference should be equal to zero,
as the first transition between neighboring Landau levels occurs
when the cyclotron and Zeeman energies are equal. Due to
electron-electron interactions this energy difference is finite as
shown in the inset of Fig.\ref{figure5} as a function of the
perpendicular magnetic field. These values represent the gain in
the exchange-correlation energy as the system jumps from spin
unpolarized to a partially spin-polarized configuration. The
values obtained are significantly large and comparable or even
larger to values obtained in GaAs- and AlAs-based systems
\cite{Daneshvar97,DePoortere00}. We see that the data points,
including those obtained at different gate voltages (different
densities), suggest a linear dependence vs $B_\perp$ similar to
what was reported in \cite{DePoortere00} and discussed in
\cite{Leadley98}. From these data we conclude that, even for
InGaAs-based systems with large In concentration, the
electron-electron interaction terms play a dominant role in
determining the magnetic transitions at low filling factors,
unlike previous studies on InAs systems characterized by larger
electron densities \cite{Brosig00}.

Finally we want to further comment on the impact of the
exchange-correlation energy gap on the magnetic transition in
tilted field. The behavior shown in Fig.\ref{figure2} and
Fig.\ref{figure4} demonstrates that while a clear crossing
characterizes the spin transitions at high even-integer filling
factors, an anomaly in the $R_{xx}$ peak positions close to the
transition region is observed at the lower filling factors $\nu=4$
and $\nu=6$. We believe that this evolution signals a partial
suppression of the quantum Hall gap at the transition. This is
consistent with the expected first-order phase transition
associated to a non-vanishing gap controlled by the exchange
interaction \cite{Giuliani85}. It is also worth considering the
impact of the Rashba coupling which can also lead to a repulsion
of the Landau levels at crossing through the mixing of the two
spin states \cite{Falko92,Falko93}. Finally it must be stressed
that at the coincidence angle the magneto-resistance $R_{xx}$
exhibits no spikes and no hysteresis at any filling factor. These
effects which were reported in a number of material systems are
associated to the formation of magnetic domains
\cite{Piazza03,DePoortere00,Jaroszynski02}. As proposed in
Ref.\cite{Jungwirth01}, the absence of resistance spikes as well
as detectable hysteresis can be explained by the presence of small
and dilute magnetic domains at the transition, which in our case
could be associated to fluctuations at short-length scale of the
In content in the quantum well.

\section{Conclusion}
In conclusion, we have reported the magneto-transport analysis of
a 2DES confined in a
In$_{0.75}$Ga$_{0.25}$As/In$_{0.75}$Al$_{0.25}$As quantum well.
Several quantum Hall states were observed and the effective
g-factor was measured in a tilted-field configuration. The bare
g-factor was estimated of the order of $\vert g_0\vert=5$, while
the impact of electron-electron interactions was shown to play a major role
at low filling factors. We have demonstrated that electron
densities down to $10^{11}$ cm$^{-2}$ can be achieved in nominally
undoped structures by applying a gate voltage. The large measured
electron mobilities at these relatively low densities make this
electron system a promising candidate for those spintronics
applications that would require high g-factors and gate-voltage
control of the electron density. In addition, this system might be of
potentially high interest for the investigation of the spin-orbit
coupling and its impact on the magnetic transition of quantum Hall
ferromagnetic states.

The authors are grateful to T. Jungwirth for helpful discussions
and for Hartree-Fock calculations. This work was supported in part
by the Ministry of University and Research (MIUR) under the FIRB
"Nanotechnologies and nanodevices for the information society" and
COFIN programs and by the European Research and Training Network COLLECT
(Project HPRN-CT-2002-00291).

\bibliography{desrat_biblio}

\begin{thebibliography}{29}
\expandafter\ifx\csname natexlab\endcsname\relax\def\natexlab#1{#1}\fi
\expandafter\ifx\csname bibnamefont\endcsname\relax
  \def\bibnamefont#1{#1}\fi
\expandafter\ifx\csname bibfnamefont\endcsname\relax
  \def\bibfnamefont#1{#1}\fi
\expandafter\ifx\csname citenamefont\endcsname\relax
  \def\citenamefont#1{#1}\fi
\expandafter\ifx\csname url\endcsname\relax
  \def\url#1{\texttt{#1}}\fi
\expandafter\ifx\csname urlprefix\endcsname\relax\def\urlprefix{URL }\fi
\providecommand{\bibinfo}[2]{#2}
\providecommand{\eprint}[2][]{\url{#2}}

\bibitem[{\citenamefont{Jakob et~al.}(2000)\citenamefont{Jakob, Stahl, Knoch,
  Appenzeller, Lengeler, Hardtdegen, and L{\"u}th}}]{Jakob00}
\bibinfo{author}{\bibfnamefont{M.}~\bibnamefont{Jakob}},
  \bibinfo{author}{\bibfnamefont{H.}~\bibnamefont{Stahl}},
  \bibinfo{author}{\bibfnamefont{J.}~\bibnamefont{Knoch}},
  \bibinfo{author}{\bibfnamefont{J.}~\bibnamefont{Appenzeller}},
  \bibinfo{author}{\bibfnamefont{B.}~\bibnamefont{Lengeler}},
  \bibinfo{author}{\bibfnamefont{H.}~\bibnamefont{Hardtdegen}},
  \bibnamefont{and} \bibinfo{author}{\bibfnamefont{H.}~\bibnamefont{L{\"u}th}},
  \bibinfo{journal}{Appl. Phys. Lett.} \textbf{\bibinfo{volume}{76}},
  \bibinfo{pages}{1152} (\bibinfo{year}{2000}).

\bibitem[{\citenamefont{Uhlisch et~al.}(2000)\citenamefont{Uhlisch, Lachenmann,
  Sch{\"a}pers, Braginsky, L{\"u}th, Appenzeller, Golubov, and
  Ustinov}}]{Uhlisch00}
\bibinfo{author}{\bibfnamefont{D.}~\bibnamefont{Uhlisch}},
  \bibinfo{author}{\bibfnamefont{S.~G.} \bibnamefont{Lachenmann}},
  \bibinfo{author}{\bibfnamefont{T.}~\bibnamefont{Sch{\"a}pers}},
  \bibinfo{author}{\bibfnamefont{A.~I.} \bibnamefont{Braginsky}},
  \bibinfo{author}{\bibfnamefont{H.}~\bibnamefont{L{\"u}th}},
  \bibinfo{author}{\bibfnamefont{J.}~\bibnamefont{Appenzeller}},
  \bibinfo{author}{\bibfnamefont{A.~A.} \bibnamefont{Golubov}},
  \bibnamefont{and} \bibinfo{author}{\bibfnamefont{A.~V.}
  \bibnamefont{Ustinov}}, \bibinfo{journal}{Phys. Rev. B}
  \textbf{\bibinfo{volume}{61}}, \bibinfo{pages}{12463} (\bibinfo{year}{2000}).

\bibitem[{\citenamefont{Dobers et~al.}(1989)\citenamefont{Dobers, Vieren,
  Guldner, Bove, Omnes, and Razeghi}}]{Dobers89}
\bibinfo{author}{\bibfnamefont{M.}~\bibnamefont{Dobers}},
  \bibinfo{author}{\bibfnamefont{J.~P.} \bibnamefont{Vieren}},
  \bibinfo{author}{\bibfnamefont{Y.}~\bibnamefont{Guldner}},
  \bibinfo{author}{\bibfnamefont{P.}~\bibnamefont{Bove}},
  \bibinfo{author}{\bibfnamefont{F.}~\bibnamefont{Omnes}}, \bibnamefont{and}
  \bibinfo{author}{\bibfnamefont{M.}~\bibnamefont{Razeghi}},
  \bibinfo{journal}{Phys. Rev. B} \textbf{\bibinfo{volume}{40}},
  \bibinfo{pages}{8075} (\bibinfo{year}{1989}).

\bibitem[{\citenamefont{Kita et~al.}(2001)\citenamefont{Kita, Sato, Gozu, and
  Yamada}}]{Kita01}
\bibinfo{author}{\bibfnamefont{T.}~\bibnamefont{Kita}},
  \bibinfo{author}{\bibfnamefont{Y.}~\bibnamefont{Sato}},
  \bibinfo{author}{\bibfnamefont{S.}~\bibnamefont{Gozu}}, \bibnamefont{and}
  \bibinfo{author}{\bibfnamefont{S.}~\bibnamefont{Yamada}},
  \bibinfo{journal}{Physica B} \textbf{\bibinfo{volume}{298}},
  \bibinfo{pages}{65} (\bibinfo{year}{2001}).

\bibitem[{\citenamefont{Gui et~al.}(2000)\citenamefont{Gui, Hu, Chen, Zheng,
  Guo, Chu, Chen, and Li}}]{Gui00}
\bibinfo{author}{\bibfnamefont{Y.~S.} \bibnamefont{Gui}},
  \bibinfo{author}{\bibfnamefont{C.~M.} \bibnamefont{Hu}},
  \bibinfo{author}{\bibfnamefont{Z.~H.} \bibnamefont{Chen}},
  \bibinfo{author}{\bibfnamefont{G.~Z.} \bibnamefont{Zheng}},
  \bibinfo{author}{\bibfnamefont{S.~L.} \bibnamefont{Guo}},
  \bibinfo{author}{\bibfnamefont{J.~H.} \bibnamefont{Chu}},
  \bibinfo{author}{\bibfnamefont{J.~X.} \bibnamefont{Chen}}, \bibnamefont{and}
  \bibinfo{author}{\bibfnamefont{A.~Z.} \bibnamefont{Li}},
  \bibinfo{journal}{Phys. Rev. B} \textbf{\bibinfo{volume}{61}},
  \bibinfo{pages}{7237} (\bibinfo{year}{2000}).

\bibitem[{\citenamefont{Gilbert and Bird}(2000)}]{Gilbert00}
\bibinfo{author}{\bibfnamefont{M.~J.} \bibnamefont{Gilbert}} \bibnamefont{and}
  \bibinfo{author}{\bibfnamefont{J.~P.} \bibnamefont{Bird}},
  \bibinfo{journal}{Appl. Phys. Lett.} \textbf{\bibinfo{volume}{77}},
  \bibinfo{pages}{1050} (\bibinfo{year}{2000}).

\bibitem[{\citenamefont{Nitta et~al.}(1997)\citenamefont{Nitta, Akazaki,
  Takayanagi, and Enoki}}]{Nitta97}
\bibinfo{author}{\bibfnamefont{J.}~\bibnamefont{Nitta}},
  \bibinfo{author}{\bibfnamefont{T.}~\bibnamefont{Akazaki}},
  \bibinfo{author}{\bibfnamefont{H.}~\bibnamefont{Takayanagi}},
  \bibnamefont{and} \bibinfo{author}{\bibfnamefont{T.}~\bibnamefont{Enoki}},
  \bibinfo{journal}{Phys. Rev. Lett.} \textbf{\bibinfo{volume}{78}},
  \bibinfo{pages}{1335} (\bibinfo{year}{1997}).

\bibitem[{\citenamefont{Sinova et~al.}(2003)\citenamefont{Sinova, Culcer, Niu,
  Sinitsyn, Jungwirth, and MacDonald}}]{Jungwirth03}
\bibinfo{author}{\bibfnamefont{J.}~\bibnamefont{Sinova}},
  \bibinfo{author}{\bibfnamefont{D.}~\bibnamefont{Culcer}},
  \bibinfo{author}{\bibfnamefont{Q.}~\bibnamefont{Niu}},
  \bibinfo{author}{\bibfnamefont{N.~A.} \bibnamefont{Sinitsyn}},
  \bibinfo{author}{\bibfnamefont{T.}~\bibnamefont{Jungwirth}},
  \bibnamefont{and} \bibinfo{author}{\bibfnamefont{A.~H.}
  \bibnamefont{MacDonald}}, \bibinfo{journal}{cond-mat/0307663}
  (\bibinfo{year}{2003}).

\bibitem[{\citenamefont{Giuliani and Quinn}(1985)}]{Giuliani85}
\bibinfo{author}{\bibfnamefont{G.~F.} \bibnamefont{Giuliani}} \bibnamefont{and}
  \bibinfo{author}{\bibfnamefont{J.~J.} \bibnamefont{Quinn}},
  \bibinfo{journal}{Phys. Rev. B} \textbf{\bibinfo{volume}{31}},
  \bibinfo{pages}{6228} (\bibinfo{year}{1985}).

\bibitem[{\citenamefont{Koch et~al.}(1993)\citenamefont{Koch, Haug,
  v.~Klitzing, and Razeghi}}]{Koch93}
\bibinfo{author}{\bibfnamefont{S.}~\bibnamefont{Koch}},
  \bibinfo{author}{\bibfnamefont{R.~J.} \bibnamefont{Haug}},
  \bibinfo{author}{\bibfnamefont{K.}~\bibnamefont{v.~Klitzing}},
  \bibnamefont{and} \bibinfo{author}{\bibfnamefont{M.}~\bibnamefont{Razeghi}},
  \bibinfo{journal}{Phys. Rev. B} \textbf{\bibinfo{volume}{47}},
  \bibinfo{pages}{4048} (\bibinfo{year}{1993}).

\bibitem[{\citenamefont{Nicholas et~al.}(1988)\citenamefont{Nicholas, Haug,
  v.~Klitzing, and Weimann}}]{Nicholas88}
\bibinfo{author}{\bibfnamefont{R.~J.} \bibnamefont{Nicholas}},
  \bibinfo{author}{\bibfnamefont{R.~J.} \bibnamefont{Haug}},
  \bibinfo{author}{\bibfnamefont{K.}~\bibnamefont{v.~Klitzing}},
  \bibnamefont{and} \bibinfo{author}{\bibfnamefont{G.}~\bibnamefont{Weimann}},
  \bibinfo{journal}{Phys. Rev. B} \textbf{\bibinfo{volume}{37}},
  \bibinfo{pages}{1294} (\bibinfo{year}{1988}).

\bibitem[{\citenamefont{Fal'ko}(1992)}]{Falko92}
\bibinfo{author}{\bibfnamefont{V.~I.} \bibnamefont{Fal'ko}},
  \bibinfo{journal}{Phys. Rev. B} \textbf{\bibinfo{volume}{46}},
  \bibinfo{pages}{4320} (\bibinfo{year}{1992}).

\bibitem[{\citenamefont{Fal'ko}(1993)}]{Falko93}
\bibinfo{author}{\bibfnamefont{V.~I.} \bibnamefont{Fal'ko}},
  \bibinfo{journal}{Phys. Rev. Lett.} \textbf{\bibinfo{volume}{71}},
  \bibinfo{pages}{141} (\bibinfo{year}{1993}).

\bibitem[{\citenamefont{Brosig et~al.}(2000)\citenamefont{Brosig, Ensslin,
  Jansen, Nguyen, Brar, Thomas, and Kroemer}}]{Brosig00}
\bibinfo{author}{\bibfnamefont{S.}~\bibnamefont{Brosig}},
  \bibinfo{author}{\bibfnamefont{K.}~\bibnamefont{Ensslin}},
  \bibinfo{author}{\bibfnamefont{A.~G.} \bibnamefont{Jansen}},
  \bibinfo{author}{\bibfnamefont{C.}~\bibnamefont{Nguyen}},
  \bibinfo{author}{\bibfnamefont{B.}~\bibnamefont{Brar}},
  \bibinfo{author}{\bibfnamefont{M.}~\bibnamefont{Thomas}}, \bibnamefont{and}
  \bibinfo{author}{\bibfnamefont{H.}~\bibnamefont{Kroemer}},
  \bibinfo{journal}{Phys. Rev. B} \textbf{\bibinfo{volume}{61}},
  \bibinfo{pages}{13045} (\bibinfo{year}{2000}).

\bibitem[{\citenamefont{Sch{\"a}pers et~al.}(1998)\citenamefont{Sch{\"a}pers,
  Engels, Lange, Klocke, Hollfelder, and L{\"u}th}}]{Schapers98}
\bibinfo{author}{\bibfnamefont{T.}~\bibnamefont{Sch{\"a}pers}},
  \bibinfo{author}{\bibfnamefont{G.}~\bibnamefont{Engels}},
  \bibinfo{author}{\bibfnamefont{J.}~\bibnamefont{Lange}},
  \bibinfo{author}{\bibfnamefont{T.}~\bibnamefont{Klocke}},
  \bibinfo{author}{\bibfnamefont{M.}~\bibnamefont{Hollfelder}},
  \bibnamefont{and} \bibinfo{author}{\bibfnamefont{H.}~\bibnamefont{L{\"u}th}},
  \bibinfo{journal}{J. Appl. Phys.} \textbf{\bibinfo{volume}{83}},
  \bibinfo{pages}{4324} (\bibinfo{year}{1998}).

\bibitem[{\citenamefont{Capotondi et~al.}(2003)\citenamefont{Capotondi,
  Biasiol, Vobornik, Sorba, Giazotto, Cavallini, and Fraboni}}]{Capotondi03}
\bibinfo{author}{\bibfnamefont{F.}~\bibnamefont{Capotondi}},
  \bibinfo{author}{\bibfnamefont{G.}~\bibnamefont{Biasiol}},
  \bibinfo{author}{\bibfnamefont{I.}~\bibnamefont{Vobornik}},
  \bibinfo{author}{\bibfnamefont{L.}~\bibnamefont{Sorba}},
  \bibinfo{author}{\bibfnamefont{F.}~\bibnamefont{Giazotto}},
  \bibinfo{author}{\bibfnamefont{A.}~\bibnamefont{Cavallini}},
  \bibnamefont{and} \bibinfo{author}{\bibfnamefont{B.}~\bibnamefont{Fraboni}},
  \bibinfo{journal}{accepted in J. Vac. Sci. Technol. B}
  (\bibinfo{year}{2003}).

\bibitem[{Sni()}]{Snider}
\bibinfo{note}{{\it 1D Poisson} by G. Snider,
  http://www.nd.edu/{$\sim$}gsnider/}.

\bibitem[{\citenamefont{Koga et~al.}(2002)\citenamefont{Koga, Nitta, Akazaki,
  and Takayanagi}}]{Koga02}
\bibinfo{author}{\bibfnamefont{T.}~\bibnamefont{Koga}},
  \bibinfo{author}{\bibfnamefont{J.}~\bibnamefont{Nitta}},
  \bibinfo{author}{\bibfnamefont{T.}~\bibnamefont{Akazaki}}, \bibnamefont{and}
  \bibinfo{author}{\bibfnamefont{H.}~\bibnamefont{Takayanagi}},
  \bibinfo{journal}{Physica E} \textbf{\bibinfo{volume}{13}},
  \bibinfo{pages}{542} (\bibinfo{year}{2002}).

\bibitem[{\citenamefont{Wang and Vasilopoulos}(2003)}]{Wang03}
\bibinfo{author}{\bibfnamefont{X.~F.} \bibnamefont{Wang}} \bibnamefont{and}
  \bibinfo{author}{\bibfnamefont{P.}~\bibnamefont{Vasilopoulos}},
  \bibinfo{journal}{Phys. Rev. B} \textbf{\bibinfo{volume}{67}},
  \bibinfo{pages}{085313} (\bibinfo{year}{2003}).

\bibitem[{\citenamefont{Ando and Uemura}(1974)}]{Ando74}
\bibinfo{author}{\bibfnamefont{T.}~\bibnamefont{Ando}} \bibnamefont{and}
  \bibinfo{author}{\bibfnamefont{Y.}~\bibnamefont{Uemura}},
  \bibinfo{journal}{J. Phys. Soc. Jpn.} \textbf{\bibinfo{volume}{37}},
  \bibinfo{pages}{1044} (\bibinfo{year}{1974}).

\bibitem[{\citenamefont{Hu et~al.}(2003)\citenamefont{Hu, Zehnder, Heyn, and
  Heitmann}}]{Hu03}
\bibinfo{author}{\bibfnamefont{C.~M.} \bibnamefont{Hu}},
  \bibinfo{author}{\bibfnamefont{C.}~\bibnamefont{Zehnder}},
  \bibinfo{author}{\bibfnamefont{C.}~\bibnamefont{Heyn}}, \bibnamefont{and}
  \bibinfo{author}{\bibfnamefont{D.}~\bibnamefont{Heitmann}},
  \bibinfo{journal}{Phys. Rev. B} \textbf{\bibinfo{volume}{67}},
  \bibinfo{pages}{201302} (\bibinfo{year}{2003}).

\bibitem[{\citenamefont{M{\"o}ller et~al.}(2003)\citenamefont{M{\"o}ller, Heyn,
  and Grundler}}]{Moller03}
\bibinfo{author}{\bibfnamefont{C.~H.} \bibnamefont{M{\"o}ller}},
  \bibinfo{author}{\bibfnamefont{C.}~\bibnamefont{Heyn}}, \bibnamefont{and}
  \bibinfo{author}{\bibfnamefont{D.}~\bibnamefont{Grundler}},
  \bibinfo{journal}{Appl. Phys. Lett.} \textbf{\bibinfo{volume}{83}},
  \bibinfo{pages}{2181} (\bibinfo{year}{2003}).

\bibitem[{\citenamefont{Nicholas et~al.}(1983)\citenamefont{Nicholas, Brummell,
  Portal, Cheng, Cho, and Pearsall}}]{Nicholas83}
\bibinfo{author}{\bibfnamefont{R.~J.} \bibnamefont{Nicholas}},
  \bibinfo{author}{\bibfnamefont{M.~A.} \bibnamefont{Brummell}},
  \bibinfo{author}{\bibfnamefont{J.~C.} \bibnamefont{Portal}},
  \bibinfo{author}{\bibfnamefont{K.~Y.} \bibnamefont{Cheng}},
  \bibinfo{author}{\bibfnamefont{A.~Y.} \bibnamefont{Cho}}, \bibnamefont{and}
  \bibinfo{author}{\bibfnamefont{T.~P.} \bibnamefont{Pearsall}},
  \bibinfo{journal}{Solid State Commun.} \textbf{\bibinfo{volume}{45}},
  \bibinfo{pages}{911} (\bibinfo{year}{1983}).

\bibitem[{\citenamefont{Daneshvar et~al.}(1997)\citenamefont{Daneshvar, Ford,
  Simmons, Khaetskii, Hamilton, Pepper, and Ritchie}}]{Daneshvar97}
\bibinfo{author}{\bibfnamefont{A.~J.} \bibnamefont{Daneshvar}},
  \bibinfo{author}{\bibfnamefont{C.~J.~B.} \bibnamefont{Ford}},
  \bibinfo{author}{\bibfnamefont{M.~Y.} \bibnamefont{Simmons}},
  \bibinfo{author}{\bibfnamefont{A.~V.} \bibnamefont{Khaetskii}},
  \bibinfo{author}{\bibfnamefont{A.~R.} \bibnamefont{Hamilton}},
  \bibinfo{author}{\bibfnamefont{M.}~\bibnamefont{Pepper}}, \bibnamefont{and}
  \bibinfo{author}{\bibfnamefont{D.~A.} \bibnamefont{Ritchie}},
  \bibinfo{journal}{Phys. Rev. Lett.} \textbf{\bibinfo{volume}{79}},
  \bibinfo{pages}{4449} (\bibinfo{year}{1997}).

\bibitem[{\citenamefont{Poortere et~al.}(2000)\citenamefont{Poortere, Tutuc,
  Papadakis, and Shayegan}}]{DePoortere00}
\bibinfo{author}{\bibfnamefont{E.~P.~D.} \bibnamefont{Poortere}},
  \bibinfo{author}{\bibfnamefont{E.}~\bibnamefont{Tutuc}},
  \bibinfo{author}{\bibfnamefont{S.~J.} \bibnamefont{Papadakis}},
  \bibnamefont{and} \bibinfo{author}{\bibfnamefont{M.}~\bibnamefont{Shayegan}},
  \bibinfo{journal}{Science} \textbf{\bibinfo{volume}{290}},
  \bibinfo{pages}{1546} (\bibinfo{year}{2000}).

\bibitem[{\citenamefont{Leadley et~al.}(1998)\citenamefont{Leadley, Nicholas,
  Harris, and Foxon}}]{Leadley98}
\bibinfo{author}{\bibfnamefont{D.~R.} \bibnamefont{Leadley}},
  \bibinfo{author}{\bibfnamefont{R.~J.} \bibnamefont{Nicholas}},
  \bibinfo{author}{\bibfnamefont{J.~J.} \bibnamefont{Harris}},
  \bibnamefont{and} \bibinfo{author}{\bibfnamefont{C.~T.} \bibnamefont{Foxon}},
  \bibinfo{journal}{Phys. Rev. B} \textbf{\bibinfo{volume}{58}},
  \bibinfo{pages}{13036} (\bibinfo{year}{1998}).

\bibitem[{\citenamefont{Piazza et~al.}(2003)\citenamefont{Piazza, Pellegrini,
  Beltram, and Wegscheider}}]{Piazza03}
\bibinfo{author}{\bibfnamefont{V.}~\bibnamefont{Piazza}},
  \bibinfo{author}{\bibfnamefont{V.}~\bibnamefont{Pellegrini}},
  \bibinfo{author}{\bibfnamefont{F.}~\bibnamefont{Beltram}}, \bibnamefont{and}
  \bibinfo{author}{\bibfnamefont{W.}~\bibnamefont{Wegscheider}},
  \bibinfo{journal}{Solid State Commun.} \textbf{\bibinfo{volume}{127}},
  \bibinfo{pages}{163} (\bibinfo{year}{2003}).

\bibitem[{\citenamefont{Jaroszy{\'n}ski
  et~al.}(2002)\citenamefont{Jaroszy{\'n}ski, Andrearczyk, Karczewski,
  Wr{\'o}bel, Wojtowicz, Papis, Kami{\'n}ska, Piotrowska, Popovi{\'c}, and
  Dietl}}]{Jaroszynski02}
\bibinfo{author}{\bibfnamefont{J.}~\bibnamefont{Jaroszy{\'n}ski}},
  \bibinfo{author}{\bibfnamefont{T.}~\bibnamefont{Andrearczyk}},
  \bibinfo{author}{\bibfnamefont{G.}~\bibnamefont{Karczewski}},
  \bibinfo{author}{\bibfnamefont{J.}~\bibnamefont{Wr{\'o}bel}},
  \bibinfo{author}{\bibfnamefont{T.}~\bibnamefont{Wojtowicz}},
  \bibinfo{author}{\bibfnamefont{E.}~\bibnamefont{Papis}},
  \bibinfo{author}{\bibfnamefont{E.}~\bibnamefont{Kami{\'n}ska}},
  \bibinfo{author}{\bibfnamefont{A.}~\bibnamefont{Piotrowska}},
  \bibinfo{author}{\bibfnamefont{D.}~\bibnamefont{Popovi{\'c}}},
  \bibnamefont{and} \bibinfo{author}{\bibfnamefont{T.}~\bibnamefont{Dietl}},
  \bibinfo{journal}{Phys. Rev. Lett.} \textbf{\bibinfo{volume}{89}},
  \bibinfo{pages}{266802} (\bibinfo{year}{2002}).

\bibitem[{\citenamefont{Jungwirth and MacDonald}(2001)}]{Jungwirth01}
\bibinfo{author}{\bibfnamefont{T.}~\bibnamefont{Jungwirth}} \bibnamefont{and}
  \bibinfo{author}{\bibfnamefont{A.~H.} \bibnamefont{MacDonald}},
  \bibinfo{journal}{Phys. Rev. Lett.} \textbf{\bibinfo{volume}{87}},
  \bibinfo{pages}{216801} (\bibinfo{year}{2001}).

\end{thebibliography}

\end{document}